\documentclass[11pt]{article}

\def\CM{\cal M}

\begin{document}

\date{}
\title{\textbf{Fedosov and Riemannian supermanifolds}}
\author{\textsc{M.~Asorey}\thanks{E-mail: asorey@saturno.unizar.es} and
\textsc{P.M.~Lavrov}\thanks{On leave of absence from Tomsk State
Pedagogical University,
634041 Tomsk, Russia; E-mail: lavrov@tspu.edu.ru}\\
\\\textit{Departamento de F\'{\i}sica Te\'{o}rica,}
\\\textit{Facultad de Ciencias Universidad de Zaragoza,}
\\\textit{50009 Zaragoza, Spain}}
\maketitle

\begin{quotation}
Generalizations of  symplectic and metric structures for   supermanifolds
 are analyzed. Two types of structures are possible according to the even/odd character
of the corresponding quadratic tensors. In the even case one has a very rich set of geometric
structures: even symplectic supermanifolds (or, equivalently, supermanifolds  with  non-degenerate Poisson 
structures), even Fedosov supermanifolds and  even  Riemannian supermanifolds. 
The existence of relations among those structures is analyzed in some details. 
In the odd case,  we show that odd  Riemannian and Fedosov supermanifolds are 
characterized by  a  scalar curvature tensor. However, odd Riemannian supermanifolds can 
only have constant curvature. 
\end{quotation}

\vspace{.5cm}

\section{Introduction}

The formulation of fundamental physical theories, either classical 
or quantum, in terms of differential geometric methods  is now well
established and presents many conceptual advantages. Probably, the most
prominent examples are  the formulation of general relativity on
Riemannian manifolds and  the geometric formulation
of gauge field theories of the fundamental forces on fiber bundles.
Another essential connection was opened by the formulation of
classical mechanics (see, for example, \cite{Ar}) -- and also
classical field theories -- on symplectic manifolds. The properties
of these manifolds are now well understood. Deformation
quantization \cite{F} was formulated in terms of symplectic manifolds
with a symmetric connection compatible with the given symplectic
structure (the so-called Fedosov manifolds \cite{fm}). The geometrical
formulation of supersymmetric field theories and the  quantization of 
general gauge theories introduced a number of applications of differential
geometry based on the concept of  supermanifold, first introduced and
studied by Berezin \cite{Ber} (see also \cite{Leites, manin}). In these
cases,  supermanifolds appear  equipped with  appropriate
symplectic structures or (and) symmetric connections. On the other hand
 the geometrical formulation of the
Batalin-Vilkovisky quantization \cite{bv} is based  on  the
so-called antisymplectic supermanifolds which are supermanifolds
equipped with an antibracket \cite{geom}. Finally,  in some approaches to
 modern gauge field theory  \cite{btgl}, flat even
Fedosov supermanifolds (in the terminology adopted here) have been
used. In summary, the geometry of manifolds and supermanifolds 
percolates all  fundamental physical theories.

The paper addresses the study of  possible extensions of  symplectic
and metric structures to supermanifolds  by means of
graded symmetric and antisymmetric second-order tensor fields of
types $(2,0)$ and $(0,2)$. In particular, we analyze the cases of
even and odd symplectic supermanifolds and  even and odd Riemannian supermanifolds. 
Symplectic supermanifolds coincide with  graded non-degenerate Poisson
supermanifolds. If, in addition, a graded
symplectic supermanifold is equipped with a symmetric connection
compatible with the symplectic structure,
it becomes a (even or odd) Fedosov supermanifold. In the even case it
can be considered as a straightforward  generalization of Fedosov manifold \cite{fm}.
The scalar curvature tensor for any such  a
Fedosov supermanifold vanishes, as for standard  Fedosov manifolds. 
We prove that a graded metric supermanifold with a compatible symmetric connection
also leads to a (even or odd) Riemannian supermanifold with a unique torsionless connection. 
The scalar curvature tensor is, in general, non trivial for both odd Riemannian 
and odd Fedosov supermanifold. 

The paper is organized as follows. In Sect.~2, we  study the basic tensor field operations 
on supermanifolds: multiplication, contraction and symmetry properties. 
In Sect.~3, we consider scalar structures
which can be used for the construction of symplectic and metric
supermanifolds.  The properties of   symmetric affine connections on supermanifolds
and  their curvature tensors are analyzed in Sect.~4.  In
Sect.~5, we introduce  the concepts of even and odd Fedosov
supermanifolds and  even and odd Riemannian supermanifolds are 
analyzed in Sect.~6.  In Sect.~7, we  summarize the main results.

We use the condensed notation suggested by DeWitt \cite{DeWitt} and
definitions and notations adopted in \cite{lr}. Derivatives with
respect to the coordinates $x^i$ are understood as acting from the
left and are  denoted by ${\partial A}/{\partial
x^i}$. Right derivatives with respect to $x^i$ are labeled by the
subscript $"r"$ or by  $A_{,i}={\partial_r A}/{\partial
x^i}$. The Grassmann parity of any quantity $A$ is denoted
by $\epsilon (A)$.
\\

\section{Tensor fields}


Let  $\CM$ be a supermanifold with dimension ${\rm dim}\ {{\CM}}=N$ and let 
 $T_p \CM$  and $T^*_p \CM$ be the tangent and cotangent spaces at a point
$p\in \CM$ respectively. We assume that each element $X\in T_p\CM$ has a certain
Grassmann parity $\epsilon(X)$ and each element $\omega\in T^*_p\CM$ has a
dual Grassmann parity
$\epsilon(\omega)$. Consider  the Cartesian
product $\Pi^n_m$ of
$T_p\CM$  and $T^*_p\CM$:
\begin{eqnarray}
\Pi^n_m=\overbrace{T^*_p\times \cdot\cdot\cdot\times
T^*_p}^{n}\times\underbrace{ T_p\times\cdot\cdot\cdot\times
T_p}_{m}.
\end{eqnarray}

Let  ${\bf T}$  be a map  ${\bf T} : \Pi^n_m\to
\Lambda$ , that maps  each element $(\omega^1,...,
\omega^n, X_1,..., X_m)\in \Pi^n_m$  into a certain supernumber
 ${\bf T}(\omega^1,...,
\omega^n, X_1,..., X_m)\in \Lambda$, where  $\Lambda$ is a Berezin
algebra, such that $\epsilon({\bf T}(\omega^1,...,
\omega^n, X_1,..., X_m))=\epsilon(T)+\epsilon(\omega^1)+...+
\epsilon(\omega^n)+
\epsilon(X_1)+...+\epsilon(X_m)$. $T$ is an even map if
 $\epsilon({T})=0$ or an odd map if $\epsilon({T})=1$.
 A map ${\bf T}$ is a  { tensor of type (n,m) and rank $n+m$ at a point $p$}, if for all
$\omega,\sigma\in T^*_p\CM$, all $X,Y\in T_p\CM$  and all $\alpha\in
\Lambda$, it satisfies the multilinear laws:
\begin{eqnarray}
\nonumber
{\bf T}(...\omega +\sigma...)&=&{\bf T}(...\omega...)+ {\bf
T}(...\sigma...),\\
\nonumber {\bf T}(...X +Y...)&=&{\bf T}(...X...)+{\bf T}(...Y...),\\
\nonumber {\bf T}(...\omega\alpha, \sigma...)&=&{\bf T}(...\omega,
\alpha\sigma...),\\
{\bf T}(...\omega\alpha, X...)&=&{\bf T}(...\omega, \alpha
X...),\\
\nonumber {\bf T}(...X\alpha, Y...)&=&{\bf T}(...X, \alpha Y...),\\
\nonumber {\bf T}(...X\alpha)&=&{\bf T}(...X) \alpha\;.
\end{eqnarray}

Let the variables $\{x^i\}, \epsilon(x^i)=\epsilon_i$ be local
coordinates on $\CM$ in the vicinity of a
point $p\in \CM$ and  $\{e_i\}$ and  $\{e^i\}$ be the corresponding coordinate
bases in the tangent space $T_p\CM$ and the cotangent space $T^*_p\CM$,
respectively. Under the change of  of local coordinates ${\bar x}^{i}={\bar
x}^i(x)$  the basis vectors in $T_p\CM$ and
$T^*_p\CM$ transform as
\begin{eqnarray}
\label{vec}
 {\bar e}_i=e_j \frac{\partial_r x^j}{\partial {\bar x}^i},
 \quad
{\bar e}^i=e^j \frac{\partial {\bar x}^i}{\partial x^j}.
\end{eqnarray}
The transformation matrices satisfy the following relations:
\begin{eqnarray}
\label{unitJ}
 \frac{\partial_r {\bar x}^i}{\partial x^k}
 \frac{\partial_r x^k}{\partial {\bar x}^j}=\delta^i_j,
 \quad
 \frac{\partial x^k}{\partial {\bar x}^j}
 \frac{\partial {\bar x}^i}{\partial x^k}=\delta^i_j,
 \quad
 \frac{\partial_r x^i}{\partial {\bar x}^k}
 \frac{\partial_r {\bar x}^k}{\partial  x^j}=\delta^i_j,
 \quad
 \frac{\partial {\bar x}^k}{\partial  x^j}
 \frac{\partial  x^i}{\partial {\bar x}^k}=\delta^i_j.
\end{eqnarray}

A tensor ${\bf T}$ can be described in local components with respect to the
chosen bases
 $\{e^i\}, \{e_i\}$:
\begin{eqnarray}
\label{comT} T^{i_1...i_n}_{\;\;\;\;\;\;\;\;\;j_1...j_m}&=&{\bf
T}(e^{i_1},...,e^{i_n}, e_{j_1},..., e_{j_m}),\\
\nonumber \epsilon(T^{i_1...i_n}_{\;\;\;\;\;\;\;\;\;\;j_1...j_m})&=&
\epsilon(T)+\epsilon_{i_1}+\cdot\cdot\cdot + \epsilon_{i_n}+
\epsilon_{j_1}+\cdot\cdot\cdot +\epsilon_{j_m},
\end{eqnarray}
Then a tensor field of type $(n,m)$ and rank $n+m$ is by definition a
geometric object that can be given by a set of functions with $n$ upper
and $m$ lower indices in each local coordinate system
$(x)=(x^1,...,x^N)$ with the following transformation laws under a change of coordinates
$x\longrightarrow {\bar x}$
\begin{eqnarray}
\label{tenzor} {\bar
T}^{i_1...i_n}_{\;\;\;\;\;\;\;\;\;\;\;j_1...j_m}&=&
T^{l_1...l_n}_{\;\;\;\;\;\;\;\;\;\;k_1...k_m} \frac{\partial_r
x^{k_m}}{\partial {\bar x}^{j_m}}\cdot\cdot\cdot \frac{\partial_r
x^{k_1}}{\partial {\bar x}^{j_1}} \frac{\partial {\bar
x}^{i_n}}{\partial x^{l_n}}\cdot\cdot\cdot
\frac{\partial {\bar x}^{i_1}}{\partial x^{l_1}}
(-1)^{P^{i_1...i_n\;k_1...k_m}_{l_1...l_n\;j_1...j_m}}\,
\end{eqnarray}
where
\begin{eqnarray}
P^{i_1...i_n\;k_1...k_m}_{l_1...l_n\;j_1...j_m}=
\sum\limits^{n}_{s=1}\sum\limits^{m}_{p=1}\epsilon_{j_p}
(\epsilon_{i_s}+\epsilon_{l_s})+
\sum\limits^{n-1}_{s=1}\sum\limits^{n}_{p=s+1}
\epsilon_{i_p}(\epsilon_{i_s}+\epsilon_{l_s})+\sum\limits^{m-1}_{s=1}
\sum\limits^{m}_{p=s+1}
\epsilon_{j_p}(\epsilon_{j_s}+\epsilon_{k_s}).
\end{eqnarray}
In the simplest case, the relations for vector fields
 $T^i$ and co-vector fields $T_i$ have the form
\begin{eqnarray}
\label{formvec}
 {\bar T}^i= T^n\frac{\partial {\bar x}^i}{\partial x^n}\,, 
 \qquad
 {\bar T}_i= T_n\frac{\partial_r x^n}{\partial{\bar x}^i}
\end{eqnarray}
and for  second-rank tensor fields of different types
Eq.(\ref{tenzor})
 become
\begin{eqnarray}
\label{formup}
{\bar T}^{ij}&=&
T^{mn}\frac{\partial {\bar x}^j}{\partial x^n}
\frac{\partial {\bar x}^i}{\partial x^m}
(-1)^{\epsilon_j(\epsilon_i+\epsilon_m)},\\
\label{form}
{\bar T}_{ij}&=&
T_{mn}\frac{\partial_r x^n}{\partial {\bar x}^j}
\frac{\partial_r x^m}{\partial {\bar x}^i}
(-1)^{\epsilon_j(\epsilon_i+\epsilon_m)},\\
\label{form1} {\bar T}^i_{\;\;j}&=& T^m_{\;\;\;n}\frac{\partial_r
x^n}{\partial {\bar x}^j} \frac{\partial {\bar x}^i}{\partial x^m}
(-1)^{\epsilon_j(\epsilon_i+\epsilon_m)}\,.
\end{eqnarray}
In definitions  (\ref{vec}),  (\ref{tenzor}) we follow the rule that places
all transformation matrices  to the right. In particular, it
means that
we adopted the following representation of elements in the tangent and
cotangent spaces
\begin{eqnarray}
\label{Xomeg}
X=X^ie_i(-1)^{\epsilon_i},\;\;\;\; \omega=\omega_ie^i.
\end{eqnarray}

Note that the unit matrix $\delta^i_j$ is related to the unit tensor
field $E^i_{\;j}$  which transforms   according to the law (\ref{form1}) as
\begin{eqnarray}
\label{unit} E^i_{\;j}=\delta^i_j.
\end{eqnarray}

From a tensor field of type $(n,m)$ and rank $n+m$ with $n\neq 0,
\;m\neq 0$, one can construct a tensor field of type $(n-1,m-1)$ and
rank $n+m-2$ by contracting an upper with a lower index by the
rules (for details, see \cite{lr}). In particular, for tensor
fields of type $(1,1)$, the contraction gives the supertrace,
\begin{eqnarray}
\label{sc1} {\rm str } T=T^i_{\;\;i}\;(-1)^{\epsilon_i}.
\end{eqnarray}
Using the multiplication operation, from two tensor fields of types
$(n,0)$ and $(0,m)$, one can construct new tensor fields of type
$(n-1,m-1)$. In particular, from a vector $U^i$ and a covector $V_i$ field
one can built a scalar field
\begin{eqnarray}
\label{sc}
 (-1)^{\epsilon_i(\epsilon(V)+1)}\;U^i\;V_i =
 (-1)^{\epsilon(U)\epsilon(V) +
 \epsilon_i\epsilon(U)}\;V_i\;U^i\,,
\end{eqnarray}
which is invariant with respect to any choice of local coordinates.
Two second-rank tensor fields $U^{ij}$ and $V_{ij}$ yield  tensor
fields
\begin{eqnarray}
\nonumber
\label{contr} &&(-1)^{(\epsilon_i+\epsilon_k)\epsilon(V)+
\epsilon_k}\;U^{ik}\;V_{kj} , \quad
(-1)^{(\epsilon_i+\epsilon_k)\epsilon(V)+
\epsilon_k(\epsilon_i+\epsilon_j+1)}\;U^{ki}V_{jk},\\
&&\\
\nonumber
&&(-1)^{(\epsilon_j+\epsilon_k)\epsilon(U)+
\epsilon_i\epsilon_j}\;V_{jk}\;U^{ki} , \quad
(-1)^{(\epsilon_j+\epsilon_k)\epsilon(U)+
\epsilon_k(\epsilon_i+\epsilon_j)+\epsilon_i\epsilon_j)}\;V_{kj}U^{ik}
\end{eqnarray}
transforming  according   (\ref{form1}). The four fields are not independent, in fact 
there are only  two independent ones due to the relations
\begin{eqnarray}
\label{contrind1}
(-1)^{(\epsilon_i+\epsilon_k)\epsilon(V)+\epsilon_k}U^{ik}V_{kj}=
(-1)^{\epsilon(U)\epsilon(V)+(\epsilon_j+\epsilon_k)\epsilon(U)+
\epsilon_k(\epsilon_i+\epsilon_j)+\epsilon_i\epsilon_j}V_{kj}U^{ik},
\end{eqnarray}
\begin{eqnarray}
\label{contrind2}
(-1)^{(\epsilon_i+\epsilon_k)\epsilon(V)+
\epsilon_k(\epsilon_i+\epsilon_j+1)}\;U^{ki}V_{jk}=
(-1)^{\epsilon(U)\epsilon(V)+(\epsilon_j+\epsilon_k)\epsilon(U)+
\epsilon_i\epsilon_j}\;V_{jk}\;U^{ki}
\end{eqnarray}

Further contractions of  indices yield the scalar field
\begin{eqnarray}
\label{contrind3} (-1)^{(\epsilon_i+\epsilon_k)(\epsilon(V)+1)}
\;U^{ik}\;V_{ki} = (-1)^{\epsilon(U)\epsilon(V)+
(\epsilon_i+\epsilon_k)\epsilon(U)}\;V_{ik}\;U^{ki}\,.
\end{eqnarray}

Later on  we shall   construct  tensor fields from
tensor fields of type $(1,2)$ and vector or co-vector tensor fields.
The corresponding rules
\begin{eqnarray}
\label{cont12}
U^lV^k_{\;\;lj}(-1)^{(\epsilon(V)+\epsilon_j+1)\epsilon_l},\quad
U_lV^l_{\;\;ij}(-1)^{\epsilon(V)\epsilon_l}
\end{eqnarray}
give $(1,1)$ and $(0,2)$ tensor fields transforming in accordance with (\ref{form1}) and
(\ref{form}) respectively.

Now recalling (\ref{unit}), (\ref{contr}), (\ref{contrind1}),
(\ref{contrind2}) and (\ref{contrind3}), the  inverse
tensor field $T_{ij}$  for a non-degenerate  second-rank tensor field
$T^{ij}$ of type $(2,0)$ should be defined via the relations
\begin{eqnarray}
\label{invers1} &&(-1)^{(\epsilon_i+\epsilon_k)\epsilon(T)+
\epsilon_k}\;T^{ik}\;T_{kj}
= \delta^i_j\,,\\
\label{invers2} &&(-1)^{(\epsilon_j+\epsilon_k)\epsilon(T)+
\epsilon_j}\;T_{jk}\;T^{ki} = \delta^i_j\,,
\\
\nonumber
 &&\epsilon(T_{ij})=\epsilon(T^{ij})=
 \epsilon(T)+\epsilon_i+\epsilon_j\,,
\end{eqnarray}
and similarly for tensor fields of type (0,2). In particular,
in the even case ($\epsilon( T)=0$) from (\ref{invers1}) and (\ref{invers2})
it follows
\begin{eqnarray}
\label{einv1}
 &&(-1)^{\epsilon_k}\;T^{ik}\;T_{kj}
= \delta^i_j\,,\\
\label{einv2}
 &&(-1)^{\epsilon_j}\;T_{jk}\;T^{ki} = \delta^i_j\,,
\end{eqnarray}
and, in its turn, in the odd case ($\epsilon( T)=1$) we have
\begin{eqnarray}
\label{oinv1}
 (-1)^{\epsilon_i}\;T^{ik}\;T_{kj}
= \delta^i_j\;,&&\\
\label{oinv2}
(-1)^{\epsilon_k}\;T_{jk}\;T^{ki}=\delta^i_j\;.&&
\end{eqnarray}
Notice that the definitions (\ref{invers1}) and (\ref{invers2})
are in agreement with
the relation (\ref{contrind3}). Indeed, contracting indices in
(\ref{invers1})
to obtain a scalar (see (\ref{sc1})) we have
\begin{eqnarray}
\nonumber
(-1)^{(\epsilon_i+\epsilon_k)(\epsilon(T)+1)}\;T^{ik}\;T_{ki}
= \delta^i_i(-1)^{\epsilon_i}=(-1)^{(\epsilon_i+\epsilon_k+1)\epsilon(T)}\;T_{ki}\;T^{ik}
\end{eqnarray}
and doing the same in (\ref{invers2}) we find
\begin{eqnarray}
\nonumber
(-1)^{(\epsilon_i+\epsilon_k)\epsilon(T)}\;T_{ik}\;T^{ki} =
\delta^i_i(-1)^{\epsilon_i}=(-1)^{(\epsilon_i+\epsilon_k)(\epsilon(T)+1)+\epsilon(T)}
\;T^{ik}\;T_{ki},
\end{eqnarray}
that proves our statement. Notice that in contrast with previous 
definitions of inverse tensor fields used in \cite{lr}, the definitions (\ref{invers1}) 
and (\ref{invers2}) lead to coincidence of left and right inverse tensor fields to a given one.

It is well known that  in the tensor calculus on manifolds,
an important role is played by symmetric and antisymmetric tensor
fields. In the supersymmetric case, supermatrices have more possible
symmetry properties (eight types, see, for example, \cite{GT}), and a
natural question is whether these properties are compatible with the tensor
transformation laws. Among the eight types of supermatrices with
possible symmetry properties there exist only two which satisfy the
tensor transformation laws. In our definition of tensor fields on
supermanifolds, only the supermatrices having the generalized
symmetry or antisymmetry properties satisfy the tensor
transformation rules. Indeed, let us consider a second-rank
supermatrix of type $(2,0)$ with the generalized symmetry
(antisymmetry) 
\begin{eqnarray}
\label{sym}
 T^{ij}=(-1)^{\epsilon_i\epsilon_j}T^{ji}\quad
 (T^{ij}=-(-1)^{\epsilon_i\epsilon_j}T^{ji}).
\end{eqnarray}
 The symmetry  is invariant under  the transformation law
(\ref{formup}),
\begin{eqnarray}
\nonumber {\bar T}^{ij}= T^{mn}\frac{\partial {\bar x}^j}{\partial
x^n} \frac{\partial {\bar x}^i}{\partial x^m}
(-1)^{\epsilon_j(\epsilon_i+\epsilon_m)}= T^{nm} \frac{\partial
{\bar x}^i}{\partial x^m} \frac{\partial {\bar x}^j}{\partial x^n}
(-1)^{\epsilon_i\epsilon_n} =(-1)^{\epsilon_i\epsilon_j}{\bar
T}^{ji}
\end{eqnarray}
and similarly for  antisymmetric tensor fields. The other six possible symmetry
types of supermatrices do not survive verification of the
compatibility with adopted tensor transformation laws. It is very
important to note that for non-degenerate symmetric and
antisymmetric tensor fields, their inverse tensor fields also have
the necessary symmetry properties. For example, if we consider a
second-rank tensor field  $T^{ij}$  with symmetric
(antisymmetric) properties
($\epsilon(T^{ij})=\epsilon(T)+\epsilon_i+\epsilon_j$), from definitions (\ref{invers1}) and
(\ref{invers2}) we can then find that the inverse tensor field also satisfies
\begin{eqnarray}
\label{syminv}
 T_{ij}=(-1)^{\epsilon(T)+\epsilon_i\epsilon_j}T_{ji}\quad
 (T_{ij}=-(-1)^{\epsilon(T)+\epsilon_i\epsilon_j}T_{ji}),\quad
 \epsilon(T_{ij})=\epsilon(T)+\epsilon_i+\epsilon_j.
\end{eqnarray}
The results mean that for any even symmetric (antisymmetric) tensor field the inverse tensor 
field is also symmetric (antisymmetric),  while for any odd symmetric (antisymmetric)
tensor field the inverse tensor field is antisymmetric (symmetric).
In what follows, we  shall use symmetric and
antisymmetric tensor fields only to construct scalar  invariant fields 
defined  on supermanifolds.

\section{Scalar Fields}

Let us analyze the most relevant scalar structures on
supermanifolds which can be defined in terms of graded symmetric and
antisymmetric tensor fields.

In general, there exist eight types of second rank tensor fields
with the required symmetry properties
\begin{eqnarray}
\label{tfant1}
&&\omega^{ij}=-(-1)^{\epsilon_i\epsilon_j}\omega^{ji},\quad
\epsilon(\omega^{ij})=
\epsilon(\omega)+\epsilon_i+\epsilon_j,\\
\label{tfsyt1}
&&\Omega^{ij}=(-1)^{\epsilon_i\epsilon_j}\Omega^{ji},\;\;\;\quad
\epsilon(\Omega^{ij})=
\epsilon(\Omega)+\epsilon_i+\epsilon_j,\\
\label{tfant2}
&&E_{ij}=-(-1)^{\epsilon_i\epsilon_j}E_{ji},\quad \epsilon(E_{ij})=
\epsilon(E)+\epsilon_i+\epsilon_j,\\
\label{tfsyt2}
&&g_{ij}=(-1)^{\epsilon_i\epsilon_j}g_{ji}, \;\;\;\;\;\;\quad
\epsilon(g_{ij})=
\epsilon(g)+\epsilon_i+\epsilon_j.
\end{eqnarray}
Using these tensor fields (\ref{tfant1})-(\ref{tfsyt2})
it is not difficult to
built eight scalar structures on a supermanifold:
\begin{eqnarray}
\label{Pst}
\{A,B\}&=&\frac{\partial_r A}{\partial x^i}(-1)^{\epsilon_i\epsilon(\omega)}
\omega^{ij}\frac{\partial B}{\partial x^j},\quad \epsilon(\{A,B\})=
\epsilon(\omega)+\epsilon(A)+\epsilon(B),\\
\label{Ant}
(A,B)&=&\frac{\partial_r A}{\partial x^i}(-1)^{\epsilon_i\epsilon(\Omega)}
\Omega^{ij}\frac{\partial B}{\partial x^j},\quad \epsilon((A,B))=
\epsilon(\Omega)+\epsilon(A)+\epsilon(B),\\
\label{E}
E&=&E_{ij}dx^j\land dx^i,\quad\quad\quad\;\;\;\;\;\;
\epsilon(E_{ij}dx^j\land dx^i)=\epsilon(E),\\
\label{g}
g&=&g_{ij}dx^j\;dx^i,\quad\quad\quad\quad\;\;\;\;\;\;
\epsilon(g_{ij}dx^j\;dx^i)=\epsilon(g),
\end{eqnarray}
where $A$ and $B$ are any superfunctions.

The bilinear operation $\{A,B\}$ (\ref{Pst}) obeys the following symmetry
property
\begin{eqnarray}
\label{Pstsym}
\{A,B\}=-(-1)^{\epsilon(\omega)+(\epsilon(A)+
\epsilon(\omega))(\epsilon(B)+\epsilon(\omega))}\{B,A\}
\end{eqnarray}
which  in the even case ($\epsilon(\omega)=0$) reduces to
\begin{eqnarray}
\label{Pstsyme}
\{A,B\}=-(-1)^{(\epsilon(A)\epsilon(B)}\{B,A\}
\end{eqnarray}
and in the odd case  ($\epsilon(\omega)=1$) to
\begin{eqnarray}
\label{Pstsymo}
\{A,B\}=(-1)^{(\epsilon(A)+1))(\epsilon(B)+1)}\{B,A\}.
\end{eqnarray}
On the other hand, the bilinear operation $(A,B)$ (\ref{Ant}) has the symmetry property
\begin{eqnarray}
\label{Antsym}
(A,B)=(-1)^{\epsilon(\omega)+(\epsilon(A)+
\epsilon(\omega))(\epsilon(B)+\epsilon(\omega))}(B,A)
\end{eqnarray}
which  in the even case ($\epsilon(\omega)=0$) reduces to
\begin{eqnarray}
\label{Antsyme}
(A,B)=(-1)^{\epsilon(A)\epsilon(B)}(B,A)
\end{eqnarray}
and in the odd case  ($\epsilon(\omega)=1$) to
\begin{eqnarray}
\label{Anttsymo}
(A,B)=-(-1)^{(\epsilon(A)+1))(\epsilon(B)+1)}(B,A).
\end{eqnarray}

Let us now try verify the Jacobi identity for both operations
$\{A,B\}$ and $(A,B)$. Indeed, we have
\begin{eqnarray}
\label{PstJI}\nonumber
&&\!\!\!\!\!\!\!\!\!\!\{A,\{B,C\}\}(-1)^{(\epsilon(A)+\epsilon(\omega))
(\epsilon(C)+\epsilon(\omega))}+
{\rm cyclic\ perms.\ }(A,B,C)\\
\nonumber
&&\!\!\!\!\!\!\!\!=(1-(-1)^{\epsilon(\omega)})\Big(
\frac{\partial_r A}{\partial x^i}\omega^{ij}
\frac{\partial^2_r B}{\partial x^k\partial x^j}
\omega^{kl}\frac{\partial C}{\partial x^l}
(-1)^{(\epsilon_i+\epsilon_k)
\epsilon(\omega)+\epsilon_j(\epsilon(B)+\epsilon_k+1)+
(\epsilon(A)+\epsilon(\omega))((\epsilon(C)+\epsilon(\omega))}\\
\nonumber
&& +{\rm cyclic\ perms.\ }(A,B,C)\Big)+
\Big(\omega^{ij}
\frac{\partial \omega^{kl}}{\partial x^j}
(-1)^{\epsilon_i(\epsilon_l+\epsilon(\omega))}+ {\rm cyclic\ perms.\ }(i,k,l)\Big)\\
\nonumber
&&
(-1)^{\epsilon(C)(\epsilon(A)+\epsilon_i+\epsilon_k)+
(\epsilon(A)+\epsilon(B)+\epsilon(C)+1)\epsilon(\omega)+
\epsilon_i(\epsilon(B)+\epsilon_k)+\epsilon_k\epsilon_l}.
\end{eqnarray}
We see that  in the even case ($\epsilon(\omega)=0$)
the bilinear operation $\{A,B\}$ satisfies
the Jacobi identity
\begin{eqnarray}
\label{PstJI}
 \{A,\{B,C\}\}(-1)^{\epsilon(A)(\epsilon(C)}+
{\rm cyclic\ perms.\ }(A,B,C)\equiv 0
\end{eqnarray}
if and only if $\omega$ satisfies 
\begin{eqnarray}
\label{PstJIw}
\omega^{ij}
\frac{\partial \omega^{kl}}{\partial x^j}
(-1)^{\epsilon_i\epsilon_l}+ {\rm cyclic\ perms.\ }(i,k,l)\equiv 0.
\end{eqnarray}
In the odd case there is no possibility of satisfying the Jacobi identity.
 
The product  $(A,B)$ associated to $\Omega$ can also satisfy the Jacobi identity
because
\begin{eqnarray}
\label{AntJI}\nonumber
&&\!\!\!\!\!\!\!\!\!\!\!\!(A,(B,C))(-1)^{(\epsilon(A)+\epsilon(\Omega))(\epsilon(C)+
\epsilon(\Omega))}+
{\rm cyclic\ perms.\ }(A,B,C)\\
\nonumber
&&\!\!\!\!\!\!\!\!=(1+(-1)^{\epsilon(\Omega)})\Big( \frac{\partial_r A}
{\partial x^i}\Omega^{ij}
\frac{\partial^2_r B}{\partial x^k\partial x^j}\Omega^{kl}
\frac{\partial C}{\partial x^l}
(-1)^{(\epsilon_i+\epsilon_k)
\epsilon(\Omega)+\epsilon_j(\epsilon(B)+\epsilon_k+1)+
(\epsilon(A)+\epsilon(\Omega))((\epsilon(C)+\epsilon(\Omega))}\\
\nonumber
&&+{\rm cyclic\ perms.\ }(A,B,C)\Big)\\
\nonumber
&&+\frac{\partial_r A}{\partial x^i}\frac{\partial_r B}{\partial x^k}
\frac{\partial_r C}{\partial x^l}\Big(\Omega^{ij}
\frac{\partial \Omega^{kl}}{\partial x^j}
(-1)^{\epsilon_i(\epsilon_l+\epsilon(\Omega))}
+ {\rm cyclic\ perms.\ }(i,k,l)\Big)\\
\nonumber
&&
(-1)^{\epsilon(C)(\epsilon(A)+\epsilon_i+\epsilon_k)+
(\epsilon(A)+\epsilon(B)+\epsilon(C)+1)\epsilon(\Omega)+
\epsilon_i(\epsilon(B)+\epsilon_k)+\epsilon_k\epsilon_l}.
\end{eqnarray}
In the odd case ($\epsilon(\Omega)=1$) the 
 Jacobi's identity can be satisfied
\begin{eqnarray}
\label{AntJI}
(A,(B,C))(-1)^{(\epsilon(A)+1)(\epsilon(C)+1)}+
{\rm cyclic\ perms.\ }(A,B,C)\equiv 0
\end{eqnarray}
if and only if  $\Omega$ satisfies
\begin{eqnarray}
\label{AntJIW}
\Omega^{ij}
\frac{\partial \Omega^{kl}}{\partial x^j}
(-1)^{\epsilon_i(\epsilon_l+1)}+ {\rm cyclic\ perms.\ }(i,k,l)\equiv 0.
\end{eqnarray}
 Therefore, when identities (\ref{PstJIw}) and
(\ref{AntJIW}) hold, one can identify $\{A,B\}$
($\epsilon(\{A,B\})=\epsilon(A)+\epsilon(B)$) and $(A,B)$
($\epsilon((A,B))=\epsilon(A)+\epsilon(B)+1$) with the Poisson
bracket and the antibracket respectively.

It is also possible to  combine the Poisson bracket associated to $\omega$ and the antibracket into the
so-called graded Poisson bracket (see, for example, \cite{BB, Bering,CarF,Bering1}). 
in the following  bilinear operation
\begin{eqnarray}
\label{Pstg}
\{A,B\}_g&=&\frac{\partial_r A}{\partial x^i}
(-1)^{\epsilon_i\epsilon(\omega)}
\omega^{ij}\frac{\partial B}{\partial x^j},\quad \epsilon(\{A,B\}_g)=
\epsilon(\omega)+\epsilon(A)+\epsilon(B).
\label{omegacom}
\end{eqnarray}
From (\ref{omegacom}) it follows the following symmetry property
\begin{eqnarray}
\label{Pstsymg}
\{A,B\}_g=-(-1)^{(\epsilon(A)+
\epsilon(\omega))(\epsilon(B)+\epsilon(\omega))}\{B,A\}_g.
\end{eqnarray}
If additionally tensor fields $\omega^{ij}$ satisfy the identities
\begin{eqnarray}
\label{PstJIwg}
\omega^{ij}
\frac{\partial \omega^{kl}}{\partial x^j}
(-1)^{\epsilon_i(\epsilon_l+\epsilon(\omega))}+ {\rm cyclic\ perms.\ }(i,k,l)\equiv 0,
\end{eqnarray}
then $\{A,B\}_g$ satisfies the Jacobi identity
\begin{eqnarray}
\label{AntJIg} \{A,\{B,C\}_g\}_g(-1)^{(\epsilon(A)+\epsilon(\omega))
(\epsilon(C)+\epsilon(\omega))}+ {\rm cyclic\ perms.\ }(A,B,C)\equiv 0
\end{eqnarray}
and plays a role of a graded Poisson bracket.

 A supermanifold $\CM$
equipped with a Poisson bracket is called a Poisson supermanifold,
$({\CM}, \{,\})$. Usually a manifold $\CM$ equipped with an non-degenerate
antibracket is called an antisymplectic supermanifold $({\CM}, (,))$ or,
sometimes, an odd Poisson supermanifold (see, for example,
\cite{CarF,Bering1}).

In Eq. (\ref{tfant2}) $E$ is any graded differential 2-form. If $E$ is closed
\begin{eqnarray}
\label{Eclos}
dE=E_{ij,k}dx^k\land dx^j\land dx^i=0
\end{eqnarray}
and non-degenerate, then it defines a graded (even or odd)
symplectic supermanifold $({\CM},E)$
\cite{Leites}.
In terms of tensor fields $E_{ij}$ the closure condition  (\ref{Eclos}) can be expressed as
\begin{eqnarray}
\label{Eclos1}
E_{ij,k}(-1)^{\epsilon_i\epsilon_k}+{\rm cyclic\ perms.\ }(i,j,k)=0, \quad E_{ij}=
-(-1)^{\epsilon_i\epsilon_j}E_{ji}
\end{eqnarray}
and in terms of inverse tensor fields $E^{ij}$ Eqs. (\ref{Eclos1})
can be rewritten in the form
\begin{eqnarray}
\label{Eclosivn}
E^{il}\frac{\partial E^{jk}}{\partial x^l}
(-1)^{\epsilon_i(\epsilon_k+\epsilon(E))}+{\rm cyclic\ perms.\ }(i,j,k)=0,\quad E^{ij}=-(-1)^{\epsilon(E)+\epsilon_i\epsilon_j}E^{ji}.
\end{eqnarray}
Identifying $E^{ij}$ with tensor fields $\omega^{ij}$ in (\ref{Pst}),
we obtain in the even case ($\epsilon(E)=0$) the Poisson bracket for which the
Jacobi identity (\ref{PstJI}) follows from (\ref{Eclosivn}). Therefore, in
the even case there is one-to-one correspondence between  non-degenerate
Poisson supermanifolds and an even symplectic supermanifolds. 
In the odd case ($\epsilon(E)=1$), if we assume  $E^{ij}=\Omega^{ij}$ in  (\ref{Ant}) 
then $E^{ij}$ defines an antibracket for which the Jacobi identity (\ref{AntJIW}) 
follows from (\ref{Eclosivn}). Therefore  antisymplectic supermanifolds can be identified 
with  odd symplectic manifolds.

If the tensor field $g_{ij}$ in (\ref{g}) is non-degenerate, one has a
graded metric that can provide the  supermanifold $\CM$  with a
graded (even or odd) metric structure, giving rise to a Riemannian supermanifold $({{\CM}},g)$.
On the other hand, the inverse tensor field $g^{ij}$ also defines a bilinear operation 
with symmetry properties (\ref{Pstsymo}) or (\ref{Antsyme}) 
but it does not satisfy  the Jacobi identity.
\\

\section{Connections in Supermanifolds}

Let us introduce a covariant derivative $\nabla$ (or an affine connection
$\Gamma$) on a supermanifold ${\cal M}$. In each local coordinate system $\{ x \}$
the covariant derivative $\nabla$ is described by its components $\nabla_i \,
(\epsilon(\nabla_i)= \epsilon_i)$, which are related to the 
components the affine connection $\Gamma$
$\Gamma^i_{\;\;jk},\; (\epsilon(\Gamma^i_{\;\;jk})=\epsilon_i+\epsilon_j+
\epsilon_k)$
by
\begin{eqnarray}
\label{Cris}
e^i\nabla_j=e^k\Gamma^i_{\;\;kj}(-1)^{\epsilon_k(\epsilon_i+1)},
\quad e_i\nabla_j=-e_k\Gamma^k_{\;\;ij}
\end{eqnarray}
where $\{e_i\}$ and $\{e^i\}$ are the associated bases of   the tangent 
$T\CM$ and cotangent $T^\ast \CM$
 spaces respectively.
The choice of factors in (\ref{Cris}) is dictated by the rules (\ref{cont12}).
From  (\ref{Xomeg}), (\ref{contrind3}) and (\ref{Cris}) it follows that the action of
covariant derivative
 on scalar, vector and co-vector tensor fields is
\begin{eqnarray}
\label{scal} T\,\nabla_i&=&T_{,i}\,
\\
\label{vector} T^i\,\nabla_j&=&T^i_{\;,j}+ T^k\Gamma^i_{\;kj}
(-1)^{\epsilon_k(\epsilon_i+1)}\,
\\
T_i\,\nabla_j&=&T_{i,j}-T_k\Gamma^k_{\;ij}\,
\end{eqnarray}
and on second-rank tensor fields of type $(2,0), (0,2)$ and $(1,1)$
\begin{eqnarray}
{T}^{ij}\,{\nabla}_k&=& {T}^{ij}_{\;\;,k} +
{T}^{il}\,\Gamma^j_{\;lk}(-1)^{\epsilon_l(\epsilon_j+1)}+
{T}^{lj}\,\Gamma^i_{\;lk}
(-1)^{\epsilon_i\epsilon_j+\epsilon_l(\epsilon_i+\epsilon_j+1)}\,\\
{T}_{ij}\,{\nabla}_k&=& {T}_{ij,k} - {T}_{il}\,\Gamma^l_{\;jk}-
{T}_{lj}\,\Gamma^l_{\;ik}
(-1)^{\epsilon_i\epsilon_j+\epsilon_l\epsilon_j}\,\\
{T}^i_{\;j}\,{\nabla}_k&=& { T}^i_{\;\;j,k} -
{T}^i_{\;l}\,\Gamma^l_{\;jk} + {T}^l_{\;j}\,\Gamma^i_{\;lk}
(-1)^{\epsilon_i\epsilon_j+\epsilon_l(\epsilon_i+\epsilon_j+1)}.
\end{eqnarray}
 Similarly, the action of the covariant derivative on a tensor field
of any rank and type is given in terms of the tensor components,
the ordinary derivatives and the connection components.

The components of the affine connection do not transform as components of a
mixed tensor
\begin{eqnarray}\label{afcon}
{\bar\Gamma}^i_{\;\;jk}= (-1)^{\epsilon_n(\epsilon_m+\epsilon_j)}
 \frac{\partial_r \bar x^i}{\partial x^l}\Gamma^l_{\;\;mn}
 \frac{\partial_r  x^m}{\partial \bar x^j}
 \frac{\partial_r  x^n}{\partial \bar x^k}
 +
 \frac{\partial_r \bar x^i}{\partial x^m}
 \frac{\partial_r^2  x^m}{\partial \bar x^j \partial \bar x^k}\,.
\end{eqnarray}
In general, the connection components $\Gamma^i_{\;jk}$ are not
 (generalized) symmetric w.r.t. the lower indices. The
obstruction to this symmetry is given by  the torsion,
\begin{eqnarray}
T^i_{\;jk} := \Gamma^i_{\;jk} -
(-1)^{\epsilon_j\epsilon_k}\Gamma^i_{\;kj}\,,
\end{eqnarray}
which also transforms as a tensor field.
 A  connection  $\nabla$ is torsionless  if  $T^i_{\;jk}=0$, i.e. if  
obeys the relation
\begin{eqnarray}
\label{Crisp} \Gamma^i_{\;jk}=
(-1)^{\epsilon_j\epsilon_k}\Gamma^i_{\;kj}.
\end{eqnarray}
 Here on, we shall consider only symmetric connections.

The curvature tensor field $R^i_{\;\;mjk}$ is defined in a
coordinate basis in terms  of the commutator of covariant
derivatives, $[\nabla_i,\nabla_j]=
\nabla_i\nabla_j-(-1)^{\epsilon_i\epsilon_j}\nabla_j\nabla_i$, whose action 
on a vector field $T^i$ is
\begin{eqnarray}
\label{Rie} T^i[\nabla_j,\nabla_k]=-(-1)^{\epsilon_m(\epsilon_i+1)}
T^mR^i_{\;\;mjk}.
\end{eqnarray}
The choice of factor in r.h.s (\ref{Rie}) is dictated by the
requirement that the contraction of tensor fields of types $(1,0)$ and
$(1,3)$ yield  a tensor field of type $(1,2)$. A straightforward
calculation yields
\begin{eqnarray}
\label{R} R^i_{\;\;mjk}=-\Gamma^i_{\;\;mj,k}+
\Gamma^i_{\;\;mk,j}(-1)^{\epsilon_j\epsilon_k}+
\Gamma^i_{\;\;jn}\Gamma^n_{\;\;mk}(-1)^{\epsilon_j\epsilon_m}-
\Gamma^i_{\;\;kn}\Gamma^n_{\;\;mj}
(-1)^{\epsilon_k(\epsilon_m+\epsilon_j)}.
\end{eqnarray}
The curvature tensor field is generalized
antisymmetric,
\begin{eqnarray}
\label{Rsym}
R^i_{\;\;mjk}=-(-1)^{\epsilon_j\epsilon_k}R^i_{\;\;mkj}\,;
\end{eqnarray}
and  satisfies  the  Jacobi identity,
\begin{eqnarray}
\label{Rjac} (-1)^{\epsilon_m\epsilon_k}R^i_{\;\;mjk}
+(-1)^{\epsilon_j\epsilon_m}R^i_{\;\;jkm}
+(-1)^{\epsilon_k\epsilon_j}R^i_{\;\;kmj}\equiv 0\,.
\end{eqnarray}
Using the  Jacobi identity for the covariant derivatives,
\begin{eqnarray}
\label{}
[\nabla_i,[\nabla_j,\nabla_k]](-1)^{\epsilon_i\epsilon_k}+
[\nabla_k,[\nabla_i,\nabla_j]](-1)^{\epsilon_k\epsilon_j}+
[\nabla_j,[\nabla_k,\nabla_i]](-1)^{\epsilon_i\epsilon_j}\equiv
0\,,
\end{eqnarray}
one obtains the  Bianchi identity,
\begin{eqnarray}
\label{BI} (-1)^{\epsilon_i\epsilon_j}R^n_{\;\;mjk;i}
+(-1)^{\epsilon_i\epsilon_k}R^n_{\;\;mij;k}
+(-1)^{\epsilon_k\epsilon_j}R^n_{\;\;mki;j}\equiv 0\,,
\end{eqnarray}
with the notation $R^n_{\;\;mjk;i}:\,=R^n_{\;\;mjk}\nabla_i$.
\\

\section{Fedosov supermanifolds}

Let us consider a symplectic supermanifold $({\CM},\omega)$, i.e. a supermanifold $\CM$  with 
a closed non-degenerate graded differential 2-form $\omega$.
A symmetric connection
$\Gamma$ (covariant derivative $\nabla$)  with components
$\Gamma^i_{\;jk}$ ($\nabla_i$) in each local coordinate system
$\{x^i\}$.  $\Gamma$ is compatible with
the symplectic structure $\omega$ if  $\omega\nabla=0$. In
local coordinates  the   compatibility condition is
\begin{eqnarray}
\label{covomiv} \omega_{ij}\nabla_k=\omega_{ij,k}-\Gamma_{ijk}+
\Gamma_{jik}(-1)^{\epsilon_i\epsilon_j}=0,\quad 
\omega_{ij}=-(-1)^{\epsilon_i\epsilon_j}\omega_{ji}
\end{eqnarray}
where we use the notation
\begin{eqnarray}
\label{G} \Gamma_{ijk}=\omega_{in}\Gamma^n_{\;\;jk},\quad
\epsilon(\Gamma_{ijk})=\epsilon(\omega)+
\epsilon_i+\epsilon_j+\epsilon_k\,.
\end{eqnarray}
 Notice that for a given symplectic structure $\omega$ there
exists a large family of connections satisfying (\ref{covomiv}). 
For instance, for any symplectic  connection we have that
\begin{eqnarray}\label{Gp}
\nonumber \Gamma_{kij}&=&\frac{1}{2}\Big(\omega_{ki,j}+
\omega_{jk,i}(-1)^{(\epsilon_i+\epsilon_k)\epsilon_j}-
\omega_{ij,k}(-1)^{(\epsilon_i+\epsilon_j)\epsilon_k}\Big)\\
&& +\Big(\Pi_{kij}+\Pi_{ikj}(-1)^{\epsilon_i\epsilon_k}-
\Pi_{jki}(-1)^{(\epsilon_i+\epsilon_k)\epsilon_j}\Big)\\
\nonumber&=&- \omega_{ij,k}(-1)^{(\epsilon_i+\epsilon_j)\epsilon_k}
+\Big(\Pi_{kij}+\Pi_{ikj}(-1)^{\epsilon_i\epsilon_k}-
\Pi_{jki}(-1)^{(\epsilon_i+\epsilon_k)\epsilon_j}\Big)\\
\nonumber &=&-
\omega_{ij,k}(-1)^{(\epsilon_i+\epsilon_j)\epsilon_k}+\Pi_{kij}+
\Big(\Pi_{ikj}(-1)^{\epsilon_i\epsilon_k}-
\Pi_{jki}(-1)^{(\epsilon_i+\epsilon_k)\epsilon_j}\Big),
\end{eqnarray}
where $\Pi_{kij}$ is the symmetric part of $\Gamma_{ijk}$:
\begin{eqnarray}
\label{Pi}\Pi_{kij}= \frac{1}{2}\Big(\Gamma_{kij}+
\Gamma_{kji}(-1)^{\epsilon_i\epsilon_j}\Big),
\end{eqnarray}
which is not a tensor field.
If the connection is also  symmetric  we have that
$\Gamma_{kij}$:
\begin{eqnarray}
\label{Gammas} \Gamma_{kij}=-
\omega_{ij,k}(-1)^{(\epsilon_i+\epsilon_j)\epsilon_k}+\Pi_{kij}.
\end{eqnarray}
The presence of $\Pi_{kij}$ in (\ref{Gammas}) is very important because if
we only had
\begin{eqnarray}
\label{Gammas1} \Gamma_{kij}=-
\omega_{ij,k}(-1)^{(\epsilon_i+\epsilon_j)\epsilon_k},
\end{eqnarray}
then  $\Gamma_{kij}$ in (\ref{Gammas1}) will   not
transform  according the  corresponding rules 
for connections (\ref{afcon}). Since $\Pi_{ijk}$ is arbitrary this shows that there is not 
 a unique symmetric connection compatible with a given
symplectic structure. In Darboux coordinates $\omega_{ij,k}=0$ \cite{agm}.

 A symplectic supermanifold $({\CM},\omega)$
equipped with a symmetric symplectic connection $\Gamma$ is called
 a Fedosov supermanifold $({\CM},\omega,\Gamma)$.

Consider now  curvature tensor $R_{ijkl}$ of a  symplectic connection
\begin{eqnarray}
\label{Rs} R_{ijkl}=\omega_{in}R^n_{\;\;jkl},\quad
\epsilon(R_{ijkl})=\epsilon(\omega)+\epsilon_i+
\epsilon_j+\epsilon_k+\epsilon_l,
\end{eqnarray}
where  $R^n_{\;\;jkl}$ is defined in  (\ref{R}).
It is obvious that
\begin{eqnarray}
\label{Rans} R_{ijkl}=-(-1)^{\epsilon_k\epsilon_l}R_{ijlk},
\end{eqnarray}
and, using (\ref{R}) and (\ref{Rjac}), one deduces the
Jacobi identity for $R_{ijkl}$,
\begin{eqnarray}
\label{Rjac0} (-1)^{\epsilon_j\epsilon_l}R_{ijkl}
+(-1)^{\epsilon_l\epsilon_k}R_{iljk}
+(-1)^{\epsilon_k\epsilon_j}R_{iklj}=0\,.
\end{eqnarray}
The curvature tensor $R_{ijkl}$ is generalized symmetric w.r.t. the
first two indices (for details, see \cite{lr,gl}),
\begin{eqnarray}
\label{Ras1} R_{ijkl}=(-1)^{\epsilon_i\epsilon_j}R_{jikl}
\end{eqnarray}
and satisfies the identity
\begin{eqnarray}
\label{Rjac2}
R_{ijkl}
+(-1)^{\epsilon_l(\epsilon_i+\epsilon_k+\epsilon_j)}R_{lijk}
+(-1)^{(\epsilon_k+\epsilon_l)(\epsilon_i+\epsilon_j)}
R_{klij}+
(-1)^{\epsilon_i(\epsilon_j+\epsilon_l+\epsilon_k)}R_{jkli}=0.
\end{eqnarray}
The last statement  is proved by using the Jacobi identity
(\ref{Rjac0}) together with a cyclic change of  indices
\cite{lr}. The identity (\ref{Rjac2}) involves components of the
curvature tensor with cyclic permutation of all indices, but the
sign factors depending on the Grassmann parities of the indices do
not follow from a cyclic permutation, as is the case, for example,
of the Jacobi identity, but are defined by the permutation of the
indices that maps a given set into the original one.

With  the curvature tensor, $R_{ijkl}$, and the inverse tensor
field $\omega^{ij}$  of the symplectic structure $\omega_{ij}$,
 one can construct the only  tensor field of type $(0,2)$,
\begin{eqnarray}
 \label{R2} K_{ij}= \omega^{kn}R_{nikj}
 (-1)^{\epsilon_i\epsilon_k+(\epsilon(\omega)+1)(\epsilon_k+\epsilon_n)}
 \;=\;R^k_{\;\;ikj}\;(-1)^{\epsilon_k(\epsilon_i+1)},\quad
 \epsilon(K_{ij})=\epsilon_i+\epsilon_j.
\end{eqnarray}
This tensor satisfies  the relations \cite{gl}
\begin{eqnarray}
\label{Rl3} [1+(-1)^{\epsilon(\omega)}](K_{ij}-(-1)^{\epsilon_i\epsilon_j}K_{ji})=0
\end{eqnarray}
and is called the Ricci tensor. In the even case this tensor is symmetric  
whereas  in the odd case there are not restrictions on its (generalized)
symmetry properties. 

Now we can define the scalar curvature tensor $K$ by the formula
\begin{eqnarray}
\label{Rsc} K=\omega^{ji}K_{ij}(-1)^{\epsilon_i+\epsilon_j}=
\omega^{ji}\omega^{kn}R_{nikj}
(-1)^{\epsilon_i+\epsilon_j+\epsilon_i\epsilon_k+
(\epsilon(\omega)+1)(\epsilon_k+\epsilon_n)}.
\end{eqnarray}
From the symmetry properties of $R_{ijkl}$, it
follows that
\begin{eqnarray}
\label{Rsc1} [1+(-1)^{\epsilon(\omega)}]K=0.
\end{eqnarray}Therefore we have proved  that as in the case of Fedosov manifolds \cite{fm} the following
proposition holds 

{\bf Proposition:} {\it  Even Fedosov  supermanifolds have vanishing scalar curvature $K$.}

However, for odd Fedosov supermanifolds this curvature  is, in general, 
not vanishing. This fact was quite recently used in Ref. \cite{BB} to  generalize
the BV formalism \cite{bv}.

Consider the Bianchi identity (\ref{BI}) in the form
\begin{eqnarray}
\label{BIFSM} 
R^n_{\;\;mij;k}
-R^n_{\;\;mik;j}(-1)^{\epsilon_k\epsilon_j}
+R^n_{\;\;mjk;i}(-1)^{\epsilon_i(\epsilon_j+\epsilon_k)}\equiv 0\,.
\end{eqnarray}
Contracting indices $i$ and $n$  with the help of (\ref{R2}) we obtain 
\begin{eqnarray}
\label{BIFSM1} 
K_{mj;k}
-K_{mk;j}(-1)^{\epsilon_k\epsilon_j}
+R^n_{\;\;mjk;n}(-1)^{\epsilon_n(\epsilon_m+\epsilon_j+\epsilon_k+1)}\equiv 0\,.
\end{eqnarray}
Now using  the relations
 \begin{eqnarray}
&& K^{i}_{\;\;j}=\omega^{ik}K_{kj}(-1)^{\epsilon_k},
K^{i}_{\;\;j;m}=\omega^{ik}K_{kj;m}(-1)^{\epsilon_k}
\\
&& 
K^{i}_{\;j;i}(-1)^{\epsilon_i(\epsilon_j+1)}= \omega^{ik}K_{kj;i}(-1)^{\epsilon_k+\epsilon_i(\epsilon_j+1)},
\end{eqnarray}
from (\ref{BIFSM1}) it follows that 
\begin{eqnarray}
\label{Riccic1}  
K_{,i}-K^{j}_{\;\;i;j}(-1)^{\epsilon_j(\epsilon_i+1)}+
\omega^{jm}R^{n}_{\;mji;n}(-1)^{\epsilon_j+\epsilon_m+
\epsilon_n(\epsilon_m+\epsilon_j+\epsilon_i+1)}\equiv 0\,.
\end{eqnarray}
Since
\begin{eqnarray}
&&\label{Ricci}{\phantom {=.}}\omega^{jm}R^{n}_{\;\;mji;n}(-1)^{\epsilon_j+\epsilon_m+
\epsilon_n(\epsilon_m+\epsilon_j+\epsilon_i+1)}\\
 &&=
\omega^{jm}\omega^{np}R_{pmji;n}
(-1)^{(\epsilon_n+\epsilon_p)\epsilon(\omega)+\epsilon_j+\epsilon_m+
\epsilon_n(\epsilon_m+\epsilon_j+\epsilon_i+1)}\\
\label{Riccic2} 
&&=\omega^{np}\omega^{jm}R_{mpji;n}(-1)^{\epsilon(\omega)+
(\epsilon_j+\epsilon_m)(\epsilon(\omega)+1)+\epsilon_j\epsilon_p+
\epsilon_n+\epsilon_p}\\
&&=K^{j}_{\;\;i;j}(-1)^{\epsilon(\omega)+
\epsilon_j(\epsilon_i+1)},
\end{eqnarray}
we have
\begin{eqnarray}
\label{Riccisc1} 
K_{,i}=[1-(-1)^{\epsilon(\omega)}]K^{j}_{\;\;i;j}
(-1)^{\epsilon_j(\epsilon_i+1)}.
\end{eqnarray}
In the odd case this implies that
\begin{eqnarray}
\label{Riccisc} 
K_{,i}=2K^{j}_{\;\;i;j}
(-1)^{\epsilon_j(\epsilon_i+1)}.
\end{eqnarray}
In the even case $K_{,i}=0$ but  the relation (\ref{Riccisc1}) does not provides 
any  new information because in this case $K=0$.
\\

\section{Riemannian supermanifolds}

Let     ${\CM}$ be a supermanifold is equipped both with a metric
structure $g$
\begin{eqnarray}
\label{gform} g=g_{ij}\;dx^j dx^i,\quad
g_{ij}=(-1)^{\epsilon_i\epsilon_j}g_{ji}, \quad
\epsilon(g_{ij})=\epsilon(g)+\epsilon_i+\epsilon_j\;,
\end{eqnarray}
and $\Delta$ a  symmetric connection  with a covariant derivative
$\nabla$  compatible with the superRiemannian metric  $g$
\begin{eqnarray}
\label{covgdiv} g_{ij}\nabla_k=g_{ij,k}-g_{im}\Delta^m_{\;\;jk}-
g_{jm}\Delta^m_{\;\;ik}(-1)^{\epsilon_i\epsilon_j}=0.
\end{eqnarray}

It is easy to show that as in the case of Riemannian geometry  there exists the unique symmetric
connection $\Delta^i_{\;jk}$ which is compatible with a given metric
structure. Indeed, repeating calculations analogous to usual
Riemannian geometry we obtain the generalization of famous Christoffel  formula
for the connection in supersymmetric case
\begin{eqnarray}\label{gDelta}
\Delta^l_{\;ki}=\frac{1}{2}g^{lj}\Big(g_{ij,k}
(-1)^{\epsilon_k\epsilon_i} +g_{jk,i}(-1)^{\epsilon_i\epsilon_j}-
g_{ki,j}(-1)^{\epsilon_k\epsilon_j}\Big)(-1)^
{\epsilon_j\epsilon_i+\epsilon_j+\epsilon(g)(\epsilon_j+\epsilon_l)}.
\end{eqnarray}
It is straightforward to show  that the symbols $\Delta^l_{\;ki}$ in
(\ref{gDelta}) are transformed according  with transformation
laws (\ref{afcon}) for connections. A metric supermanifold $({{\CM}},g)$
equipped with a (even or odd) symmetric connection $\Delta$
compatible with a given metric structure $g$ can be refereed as a
(even or odd) Riemannian supermanifold $({{\CM}},g,\Delta)$.

The curvature tensor of the connection is
(\ref{gDelta})
\begin{eqnarray}
\label{Rs} {\cal R}_{ijkl}=g_{in}{\cal R}^n_{\;jkl},\quad
\epsilon({\cal R}_{ijkl})=\epsilon(g)+\epsilon_i+
\epsilon_j+\epsilon_k+\epsilon_l,
\end{eqnarray}
where ${\cal R}^n_{\;\;jkl}$ is given by (\ref{R}) with natural replacement
 $\Gamma^i_{\;jk}$ for $\Delta^i_{\;jk}$, which leads to the following representation,
\begin{eqnarray}\label{r1}\nonumber
{\cal
R}_{ijkl}&=&-\Delta_{ijk,l}+\Delta_{ijl,k}(-1)^{\epsilon_k\epsilon_l}
+\Delta_{nil}\Delta^n_{\;jk}(-1)^
{\epsilon_i\epsilon_n+\epsilon_l(\epsilon_n+\epsilon_j+\epsilon_k)}\\
&&-\Delta_{nik}\Delta^n_{jl}(-1)^{\epsilon_i\epsilon_n+\epsilon_k
(\epsilon_n+\epsilon_j)}
\end{eqnarray}
where
\begin{eqnarray}\label{D1}
\Delta_{ijk} =g_{in}\Delta^n_{\;jk}\;,\quad\epsilon(\Delta_{ijk})
=\epsilon(g)+\epsilon_i+\epsilon_j+\epsilon_k.
\end{eqnarray}
In this representation the relations (\ref{covgdiv}) and (\ref{gDelta}) read
\begin{eqnarray}
\label{Omegainvcon1} g_{ij,k}=\Delta_{ijk}+
\Delta_{jik}(-1)^{\epsilon_i\epsilon_j}\;,
\end{eqnarray}
\begin{eqnarray}
\label{Delta2} \Delta_{ijk}=-\frac{1}{2}\Big(g_{ki,j}
(-1)^{\epsilon_k\epsilon_j} +g_{ij,k}(-1)^{\epsilon_i\epsilon_k}-
g_{jk,i}(-1)^{\epsilon_i\epsilon_j}\Big)(-1)^
{\epsilon_i\epsilon_k}.
\end{eqnarray}
Furthermore, from Eq.~(\ref{r1}) it is follows that
\begin{eqnarray}
\label{Rans1} {\cal R}_{ijkl}=-(-1)^{\epsilon_k\epsilon_l}{\cal R}_{ijlk},
\end{eqnarray}
and, using (\ref{Rs}) and (\ref{Rjac}), one obtains  the
Jacobi identity for ${\cal R}_{ijkl}$,
\begin{eqnarray}
\label{Rjac1} (-1)^{\epsilon_j\epsilon_l}{\cal R}_{ijkl}
+(-1)^{\epsilon_l\epsilon_k}{\cal R}_{iljk}
+(-1)^{\epsilon_k\epsilon_j}{\cal R}_{iklj}=0\,.
\end{eqnarray}
In addition, the curvature tensor ${\cal R}_{ijkl}$ is generalized
antisymmetric w.r.t. the first two indices,
\begin{eqnarray}
\label{Ras}
{\cal R}_{ijkl}=-(-1)^{\epsilon_i\epsilon_j}{\cal R}_{jikl}.
\end{eqnarray}
In order to prove this, let us consider
\begin{eqnarray}
\label{com12} g_{ij,kl}=\Delta_{ijk,l}+
\Delta_{jik,l}(-1)^{\epsilon_i\epsilon_j}.
\end{eqnarray}
Then, using the relations
\begin{eqnarray}
\label{G1} \Delta_{ijk,l}=g_{in}\Delta^n_{\;jk,l}
+g_{in,l}\Delta^n_{\;jk}
(-1)^{(\epsilon_n+\epsilon_j+\epsilon_k)\epsilon_l}
\end{eqnarray}
and the definitions (\ref{r1}) and (\ref{com12}), we get
\begin{eqnarray}
\label{com1}
\nonumber
0&=&g_{ij,kl}-(-1)^{\epsilon_k\epsilon_l}g_{ij,lk}\\
\nonumber
&=&\Delta_{ijk,l}+
\Delta_{jik,l}(-1)^{\epsilon_i\epsilon_j}
-\Delta_{ijl,k}(-1)^{\epsilon_k\epsilon_l}-
\Delta_{jil,k}(-1)^{\epsilon_i\epsilon_j+\epsilon_k\epsilon_l}\\
&=&-{\cal R}_{ijkl}-(-1)^{\epsilon_i\epsilon_j}{\cal R}_{jikl}.
\end{eqnarray}
The tensor ${\cal R}_{ijkl}$ obeys the generalized symmetry property
w.r.t permutation of pair indices
\begin{eqnarray}
\label{Rperm}
{\cal R}_{ijkl}={\cal R}_{klij}(-1)^{(\epsilon_i+\epsilon_j)
(\epsilon_k+\epsilon_l)}.
\end{eqnarray}

Indeed, from the Jacobi identity (\ref{Rjac1}) one has
\begin{equation}\label{ri}
{\cal R}_{ijkl}(-1)^{\epsilon_j\epsilon_l}+{\cal
R}_{iljk}(-1)^{\epsilon_k\epsilon_l}+{\cal
R}_{iklj}(-1)^{\epsilon_j\epsilon_k}=0\;,
\end{equation}
\begin{equation}\label{rj}
{\cal R}_{jikl}(-1)^{\epsilon_i\epsilon_l}+{\cal
R}_{jkli}(-1)^{\epsilon_k\epsilon_i}+{\cal
R}_{jlik}(-1)^{\epsilon_l\epsilon_k}=0
\end{equation}
Now, multiplying relation (\ref{ri}) by
$(-1)^{\epsilon_i(\epsilon_j+\epsilon_l)}$,
multiplying Eq. (\ref{rj}) by $(-1)^{\epsilon_j\epsilon_l}$,
 and subtracting the results, we obtain
\begin{eqnarray}\label{2ri}
2{\cal
R}_{ijkl}(-1)^{\epsilon_j\epsilon_l+\epsilon_{i}
(\epsilon_l+\epsilon_j)}&+&{\cal
R}_{iljk}(-1)^{\epsilon_k\epsilon_l+\epsilon_{i}
(\epsilon_l+\epsilon_j)}+{\cal
R}_{iklj}(-1)^{\epsilon_j\epsilon_k+\epsilon_{i}
(\epsilon_l+\epsilon_j)}\\\nonumber
&&-{\cal
R}_{jkli}(-1)^{\epsilon_k\epsilon_i+\epsilon_j\epsilon_l}+{\cal
R}_{jlik}(-1)^{\epsilon_l\epsilon_k+\epsilon_j\epsilon_l}=0\;.
\end{eqnarray}
and by
the permutation of pair indices
\begin{eqnarray}\label{2rk}
2{\cal
R}_{klij}(-1)^{\epsilon_j\epsilon_l+\epsilon_{k}(\epsilon_l+\epsilon_j)}&
+&{\cal
R}_{kijl}(-1)^{\epsilon_i\epsilon_l+\epsilon_{k}(\epsilon_l+\epsilon_j)}
+{\cal
R}_{kjli}(-1)^{\epsilon_i\epsilon_j+\epsilon_{k}(\epsilon_l+\epsilon_j)}
\\\nonumber
&&-{\cal
R}_{lijk}(-1)^{\epsilon_k\epsilon_i+\epsilon_j\epsilon_l}+{\cal
R}_{ljki}(-1)^{\epsilon_i\epsilon_j+\epsilon_j\epsilon_l}=0.
\end{eqnarray}
Multiplying the relation (\ref{2ri}) by
$(-1)^{(\epsilon_l+\epsilon_i)(\epsilon_j+\epsilon_k)}$, subtracting
the results and using
$${\cal R}_{ijkl}=-{\cal
R}_{jikl}(-1)^{\epsilon_i\epsilon_j}\;\; , \;\;{\cal
R}_{ijkl}=-{\cal R}_{ijlk}(-1)^{\epsilon_k\epsilon_l}\;,
$$
we obtain the property (\ref{Rperm}).

From the  curvature tensor ${\cal R}_{ijkl}$ and the inverse tensor field $g^{ij}$ of the metric $g_{ij}$
\begin{eqnarray}
\label{ginv}
g^{ij}=(-1)^{\epsilon(g)+\epsilon_i\epsilon_j}g^{ji},\quad 
\epsilon(g^{ij})=\epsilon(g)+\epsilon_i+\epsilon_j,
\end{eqnarray}
one can define the following three tensor field of type $(0,2)$:
\begin{eqnarray}
\label{3tf1}
&&K_{ij}={\cal R}^k_{\;\;kij}(-1)^{\epsilon_k}=g^{kn}{\cal R}_{nkij}
(-1)^{(\epsilon_k+\epsilon_n)(\epsilon(g)+1)},\\
\label{3tf2}
&&{\cal R}_{ij}={\cal R}^k_{\;\;ikj}(-1)^{\epsilon_k(\epsilon_i+1)}=
g^{kn}{\cal R}_{nikj}
(-1)^{(\epsilon_k+\epsilon_n)(\epsilon(g)+1)+\epsilon_i\epsilon_k},\\
\label{3tf3}
&&Q_{ij}={\cal R}^k_{\;\;ijk}(-1)^{\epsilon_k(\epsilon_i+\epsilon_j+1)}=
g^{kn}{\cal R}_{nijk}
(-1)^{(\epsilon_k+\epsilon_n)(\epsilon(g)+1)+\epsilon_i\epsilon_k+\epsilon_j\epsilon_k}\\\nonumber
&&\epsilon(K_{ij})=\epsilon({\cal R}_{ij})=\epsilon(Q_{ij})=\epsilon_i+\epsilon_j.
\end{eqnarray}
Taking into account the definitions (\ref{3tf1})-(\ref{3tf3}) and the symmetry properties
(\ref{Rans1}), (\ref{Ras}), (\ref{Rperm}) and (\ref{ginv}), one can easily find the 
symmetry properties of $K_{ij}$, ${\cal R}_{ij}$ and $Q_{ij}$
\begin{eqnarray}
\label{sp2r}
K_{ij}=-(-1)^{\epsilon_i\epsilon_j}K_{ji},\quad 
{\cal R}_{ij}=(-1)^{\epsilon(g)+\epsilon_i\epsilon_j}{\cal R}_{ji},\quad
Q_{ij}=(-1)^{\epsilon(g)+\epsilon_i\epsilon_j}Q_{ji}.
\end{eqnarray}
Moreover from (\ref{3tf1}) and (\ref{Ras}), (\ref{ginv}) it follows
\begin{eqnarray}
\label{Ksp}
[1+(-1)^{\epsilon(g)}]K_{ij}=0.
\end{eqnarray}
This implies that in the even case ($\epsilon(g)=0$) $K_{ij}=0$. In a similar way one obtains 
\begin{eqnarray}
\label{QRsp}
Q_{ij}=-{\cal R}_{ij}.
\end{eqnarray}
From the Jacobi identity (\ref{Rjac1}) the relations among tensors 
$K_{ij},{\cal R}_{ij},Q_{ij}$ can be obtained
\begin{eqnarray}
\label{QRsp}
{\cal R}_{ij}+K_{ji}(-1)^{\epsilon_i\epsilon_j}+Q_{ji}(-1)^{\epsilon_i\epsilon_j}=0.
\end{eqnarray}
Therefore 
\begin{eqnarray}
\label{KRsp}
K_{ij}=[1-(-1)^{\epsilon(g)}]{\cal R}_{ij}
\end{eqnarray}
and $R_{ij}$ is  the only independent second-rank tensor field
which can be constructed from the curvature tensor ${\cal R}^i_{\;\;jkl}$. It is the generalized   Ricci tensor. 

A further contraction between the metric and Ricci tensors
define  the scalar curvature
\begin{eqnarray}
\label{Scalcur} {\cal R} = g^{ji}{\cal
R}_{ij}\;(-1)^{\epsilon_i+\epsilon_j},\quad \epsilon({\cal
R})=\epsilon(g)
\end{eqnarray}
which, in general, is not equal to zero. Notice that for an odd
metric structure the scalar curvature tensor squared is identically
equal to zero, ${\cal R}^2=0$.

Consider now relations which follow from the Bianchi identity (\ref{BI}).
Repeating all arguments given in the end of previous Section one can derive  the following 
relation between the  scalar curvature  and the Ricci tensor
\begin{eqnarray}
\label{RicciscR1} 
{\cal R}_{,i}=[1+(-1)^{\epsilon(g)}]{\cal R}^{j}_{\;\;i;j}
(-1)^{\epsilon_j(\epsilon_i+1)}.
\end{eqnarray}
In the even case we have
\begin{eqnarray}
\label{Riccisc} 
{\cal R}_{,i}=2{\cal R}^{j}_{\;\;i;j}
(-1)^{\epsilon_j(\epsilon_i+1)},
\end{eqnarray}
which is a supersymmetric generalization of known relation in Riemannian geometry \cite{Eisenhart}.
In the odd case ${\cal R}_{,i}=0$ and the relation (\ref{RicciscR1}) implies that  ${\cal R}=$const. 

Therefore we have proved the following proposition.

{\bf Proposition:} {\it  Odd Riemann supermanifolds have constant scalar curvature ${\cal R}=$ const.}
\\

\section{Discussion}

We have analyzed the natural geometric structures of  supermanifolds defined
symmetric and antisymmetric graded tensor fields of the second rank
and types $(2,0)$ and $(0,2)$. It was shown that a Poisson bracket
can be associated with an antisymmetric even tensor field of type
$(2,0)$ while an antibracket is related to an symmetrical odd tensor
field of type $(2,0)$.  We have have shown
that  all properties and relations for both even
and  odd  symplectic supermanifolds  equipped with a symmetric
connection compatible with a given symplectic structure ( even and
 odd Fedosov supermanifolds) have a similar  form.  In a similar way  both  
even and  odd metric
supermanifolds equipped with a (unique) symmetric connection
compatible with a given metric structure (even and odd Riemannian
supermanifolds) have the same algebraic properties except that in
the odd case a scalar curvature tensor squared is identically equal to zero
and the Ricci tensor is antisymmetric. It was shown that an antisymplectic 
supermanifold underlying the Batalin-Vilkovisky quantization method in general 
coordinates is just an odd Fedosov supermanifold.
It was proven that in the odd case the scalar curvature tensor for both 
Riemannian and Fedosov supermanifolds is, in general, non-zero. Odd 
Riemannian supermanifolds are however strongly constrained by the fact that
their scalar curvature has to be constant.

\section*{Acknowledgements}
The work of M.A. is partially supported by CICYT (grant FPA2006-2315)
and DGIID-DGA (grant2007-E24/2). P.M.L. acknowledges the MEC for the 
grant (SAB2006-0153). The work of P.M.L. was supported by the RFBR 
grant, project No.\
06-02-16346, the DFG grant, project No.\ 436 RUS 113/669/0-3, the
joint RFBR-DFG grant, project No.\ 06-02-04012 and the grant for
LRSS, project No.\ 4489.2006.2.


\begin{thebibliography}{99}

\bibitem{Ar}
V.I. Arnold, {\it Mathematical methods of classical mechanics},
Springer-Verlag, Berlin, Heidelberg, 1978.

\bibitem{F}
 B.V. Fedosov, {\it J. Diff. Geom.}, {\bf 40}, 213 (1994);
 {\it Deformation quantization and index theory}, Akademie
 Verlag, Berlin, 1996.

\bibitem{fm}
 I.~Gelfand, V.~Retakh and M.~Shubin, {\it Adv. Math.}, {\bf 136}, 104
(1998); [dg-ga/9707024].



\bibitem{Ber}
F.A. Berezin, {\it Yad. Fiz.}, {\bf 29}, 1670 (1979);
{\it ibid.} {\bf 30}, 1168 (1979);
{\it Introduction to superanalysis},
Reidel, Dordrecht, 1987.

\bibitem{Leites}
D.A. Leites, {\it Theory of Supermanifolds}, Petrozavodsk, 1983 (in
Russian).

\bibitem{manin} Yu. I. Manin, {\it Gauge Field Theory and Complex Geometry},
Springer-Verlag, Berlin-Heidelberg,1988

\bibitem{bv}
I.A. Batalin and G.A. Vilkovisky,
{\it Phys. Lett.}, {\bf B102}, 27 (1981);
{\it Phys. Rev.}, {\bf D28}, 2567 (1983).

\bibitem{geom}
E. Witten, {\it Mod. Phys. Lett.}, {\bf A5}, 487 (1990);
O.M. Khudaverdian, {\it J. Math. Phys.}, {\bf 32}, 1934 (1991)

\bibitem{btgl}
I.A. Batalin and I.V. Tyutin, {\it Nucl. Phys.}, {\bf B345}, 645 (1990);
B. Geyer and P.M. Lavrov, {\it Int. J. Mod. Phys.}, {\bf A19}, 1639 (2004).

\bibitem{DeWitt}
B. DeWitt,{\it Theory of groups and fields}, Gordon and Breach, New York, 1965.



\bibitem{lr}
P.M. Lavrov and O.V. Radchenko, {\it Theor. Math. Phys.}, {\bf 149},
1474 (2006); {\it Symplectic geometries on supermanifolds},
ArXiv:0708.3778 [hep-th].

\bibitem{GT}
D.M.  Gitman and I.V. Tyutin, {\it Quantization of fields with
constraints}, Springer-Verlag, 1990.





\bibitem{BB}
I.A. Batalin and K. Bering,  Odd Scalar Curvature in
Field-Antifield Formalism, arXiv:0708.0400; Odd Scalar Curvature in Anti-Poisson 
Geometry, arXiv:0712.3699.

\bibitem{Bering}
 K. Bering,
{\it Almost Parity Structure, Connections and Vielbeins in BV
Geometry}, arXiv:physics/9711010.

\bibitem{CarF}
J. F. Cari\~{n}ena and H. Figueroa, {\it Diff. Geom. Appl.}, {\bf
10} , 191 (1999).

\bibitem{Bering1}
 K. Bering, Semidensities, Second-Class Constraints and Convension in Anti-Poisson Geometry, ArXiv:0705.3440 [hep-th].


\bibitem{agm} M. Asorey, D.
Garc{\'{\i}}a-Alvarez and J.M. Mu\~noz-Casta\~neda,
 Monog. RSME {\bf 8}, 73-84 (2006)


\bibitem{gl}
B. Geyer and P.M. Lavrov, {\it Int. J. Mod. Phys.}, {\bf A19}, 3195 (2004).


\bibitem{Eisenhart} 
L.P. Eisenhart,
{\it Riemannian Geometry}, Princeton University Press, 1949.

\end{thebibliography}
\end{document}